\documentclass[reprint,
superscriptaddress,
amsmath,amssymb,aps,showkeys,showpacs,
twoside,final,secnumarabic,
nofootinbib]{revtex4-2}

\usepackage[paperwidth=205mm,paperheight=290mm,top=17mm,bottom=25mm,
inner=17mm,outer=17mm,
twoside]{geometry}

\usepackage{cmap} 
\usepackage[T1,T2A]{fontenc}
\usepackage[utf8]{inputenc}
\usepackage[russian,english]{babel}
\usepackage{color}
\usepackage{graphicx}
\usepackage{dcolumn}
\usepackage{bm} 
\usepackage[unicode=true,colorlinks=true,linkcolor=magenta, urlcolor=blue, citecolor = blue,breaklinks]{hyperref}
\usepackage{multirow}
\usepackage{url}
\usepackage{breakurl}
\DeclareGraphicsExtensions{.eps}

\newcount\issue
\newcount\Vol
\newcount\numb
\usepackage{fancyhdr} 
\def\Vol{\textbf{78}}
\def\numb{x}
\begin{document}

\title{JOURNAL SECTION OR CONFERENCE SECTION\\[20pt]
 Semileptonic B-decays \\ at Belle and Belle II} 

\def\addressa{National Research University Higher School of Economics, Russian Federation}
\def\addressb{address 2}

\makeatletter
\def\andname{}
\def\@andcomma{} 
\makeatother

\author{\firstname{N.}~\surname{Peters}}
\email[E-mail: ]{npeters@hse.ru}
\affiliation{\addressa}

\author{\\ on behalf of Belle II collaboration}

\received{xx.xx.2025}
\revised{xx.xx.2025}
\accepted{xx.xx.2025}

\begin{abstract}
The paper summarizes recent results from the Belle and Belle II collaborations on semileptonic $B$ decays measurements including inclusive and exclusive determination of Cabibbo-Kobayashi-Maskawa matrix elements $|V_{cb}|$ and $|V_{ub}|$ and lepton flavor universality tests studies. The results are based on the full Belle data and 361 fb$^{-1}$ Belle II data samples collected in 2018 -- 2022.
\end{abstract}

\par
\keywords{$B$ mesons, Semileptonic decays, Belle, Belle II, $e^+ e^-$ collisions  \\[5pt]}

\maketitle
\thispagestyle{fancy}


\section{Motivation}\label{intro}

\subsection{\label{sec:ckm_matrix} Cabibbo-Kobayashi-Maskawa matrix measurements}

In the Standard Model (SM), the coupling of different quark flavors in weak charged currents is parameterized by the Cabibbo-Kobayashi-Maskawa (CKM) matrix \cite{Kobayashi:1973fv}. Accurate measurements of matrix elements are crucial to test the CKM unitarity triangle and serve as key inputs for precise SM predictions. 

Semileptonic decays can be powerful method of research due to reduction of theoretical calculations as only part of particles in final state are taking part in strong interactions. There are two fundamentally different approaches: the study of inclusive and exclusive decays. The exclusive method focuses on measuring the rates and properties of specific decay modes, such as $B \to D^{(*)}\ell \nu$, whereas the inclusive method analyzes all possible semileptonic $B$ decays by summing over all hadronic final states. 

The inclusive and exclusive analyses are based on different theoretical models and experimental methods,
yielding measurements that differ by approximately $3\sigma$
\cite{HeavyFlavorAveragingGroupHFLAV:2024ctg} motivating further research.

\subsection{\label{sec:rd_motivation} Lepton flavor universality tests}
In the SM the $W$ boson couples equally to all generations of leptons. The corresponding symmetry is called lepton flavour universality (LFU). Consequently, observation of it's violation would become a clear signature of New Physics beyond the SM. Semileptonic $B$ meson decays can provide stringent test of LFU via the measurement of branching fractions' ratio:
\begin{equation}
\label{Rdef}
R(X_{\tau/\ell}) = \frac{\mathcal{B}(B \to X \tau \nu)}{\mathcal{B}(B \to X \ell \nu)}, 
\end{equation}
where $\ell$ denotes charged meson ($e$ or $\mu$) and $X$ represents hadron such as $D$ or $D^*$ meson or more inclusive hadronic system. Measurement of ratio leads to partial cancellation of theoretical and experimental uncertainties. 

With respect to the most recent theory calculations for $R(D^{(*)})$, the combined tension with the SM expectation persists at 3.14$\sigma$ \cite{HeavyFlavorAveragingGroupHFLAV:2024ctg}. Therefore, more measurements with high precision are needed. 

\section{\label{sec:level1} Kinematics at the \texorpdfstring{$B$}{B} factories}
The key feature of the Belle \cite{Belle:2000cnh} (Belle II \cite{Belle-II:2010dht}) located at the KEKB \cite{KEKTsukuba:1995urg} (SuperKEKB \cite{Akai:2018mbz}) accelerator is that experiments collected data at a center-of-mass energy corresponding to the $\Upsilon(4S)$ resonance decaying to $B\bar{B}$ pair.

Established approach is the usage of tagging -- the idea of constraining allowed decay chains and signal $B$ meson parameters by reconstructing second $B$ meson in event. This method allows to obtain all crucial details about decays (event type, decay vertex) and to constrain four-momentum of signal meson. There are three main approaches: hadronic, semileptonic and inclusive tagging. The first method provides fully reconstructed meson, but suffers from a low efficiency due to tiny branching fractions, whereas the semileptonic method achieves higher statistics, but miss kinematic information due to the undetected neutrino. Inclusive tagging provides the highest number of events with the lowest purity. Corresponding software tool is called Full Event Interpretation (FEI) \cite{Keck:2018lcd}.

Additionally, the mass of the $\Upsilon(4S)$ resonance is just above the $B\bar{B}$ production threshold. Therefore, $B$ mesons are produced almost at rest in c.m. system producing an isotropic distribution of the events. While processes with light quarks production $e^+e^- \to q\bar{q}, $ $q = u, d, s, c$ has a jet-like shapes. That allows to discriminate signal and background events by using topology and sphericity-related variables.  

\section{Recent results from Belle and Belle II}
\subsection{\texorpdfstring{$\mathcal{R}(D^{(*)+})$}{R(D(*)+)} with semileptonic \texorpdfstring{$B$}{B} tagging}

In this analysis $R(D^{(*)+}_{\tau/\ell})$ defined by Eq. (\ref{Rdef}) is measured \cite{Belle-II:2025yjp}. Neutral mode $\Upsilon(4S) \to B^0 \bar{B}^0$ is studied with reconstruction of one of the $B$ mesons via semileptonic FEI. Signal candidates are reconstructed from combination of $D^+$ and $D^{*+}$ with oppositely charged $\mu$ or $e$. Charged lepton can originate from either semitauonic decays $\bar{B}^0 \to D^{(*)+} \tau^- \bar{\nu}_{\tau}$ with $\tau^- \to \ell^- \bar{\nu}_{\ell}\nu_{\tau}$ or semileptonic decays $\bar{B}^0 \to D^{(*)+} \ell^- \bar{\nu}_{\ell}$. These processes are distinguishable through kinematic properties due to different number of neutrinos. To suppress background cosine of the angle between the $B$ meson’s momentum and its visible decay products in the c.m. frame is used:
\begin{equation}
    \label{cosine}
    \cos \theta_{B Y} =\frac{2 E_{\text {Beam }} E_Y-m_B^2-m_Y^2}{2\left|\vec{p}_B\right|\left|\vec{p}_Y\right|},
\end{equation}
where $E_{\text{beam}}$ is the beam energy, $m_{B}$ and $|\bar{p}_{B}|$ is the $B$ meson mass and momentum. Here $Y = D^{(*)}\ell$ represents the system of visible decay products. Due to resolution effects the selection criteria for this observable are $\cos\theta_{BY}^{\operatorname{tag}}\in[-1.75, 1.1]$ and $\cos\theta_{BY}^{\operatorname{sig}}\in[-15, 1.1]$ for tag and signal $B$ meson respectively. Additionally, multivariate classifiers trained on the kinematic properties of both $B$ mesons for distinguishing signal candidates originating from semitauonic, semileptonic, and background sources are employed.

Finally, two-dimensional binned log-likelihood fit to the output score of the signal BDT and the difference of the output scores of the normalization and background events is performed for extraction of the corresponding yields. 

Obtained results are 
\begin{eqnarray*}
R\!\left(D_{\tau / \ell}^{*+}\right) &= 0.306^{+0.035}_{-0.033} \text{ (stat)}^{+0.016}_{-0.018} \text{ (syst)}, \\
R\!\left(D_{\tau /\ell}^{+}\right) &= 0.418^{+0.075}_{-0.073} \text{ (stat)}^{+0.049}_{-0.056} \text{ (syst)}.
\end{eqnarray*}

These results are compatible with SM within $1.7\sigma$ and are in agreement with world average values. 

Extra measurement of the ratio of the semileptonic signal branching fractions of electrons to muons for another LFU test is performed:

\begin{eqnarray*}
R\!\left(D_{e / \mu}^{*+}\right) &= 1.068^{+0.037}_{-0.036} \text{ (stat)}^{+0.020}_{-0.019} \text{ (syst)}, \\
R\!\left(D_{e / \mu}^{+}\right) &= 1.079^{+0.037}_{-0.036} \text{ (stat)}^{+0.020}_{-0.019} \text{ (syst)}.
\end{eqnarray*}

These results are also consistent with the SM within $1.6 \sigma$ and $1.2 \sigma$ respectively. 

\subsection{\texorpdfstring{$|V_{ub}|/|V_{cb}|$}{|Vub|/|Vcb|} measurement from tagged \texorpdfstring{$\bar{B} \to X \ell\bar{\nu}_{\ell}$}{Bbar -> X l nubar} decays at Belle}

This analysis presents a measurement of the inclusive semileptonic decays $\bar{B} \to X_c \ell\bar{\nu}_{\ell}$ and $\bar{B} \to X_u \ell\bar{\nu}_{\ell}$ partial branching fractions ratio using the full Belle dataset collected at $\Upsilon(4S)$ resonance with $772 \times 10^6$ $B\bar{B}$ pairs \cite{Belle:2023asa}. Main analysis method is the implementation of improved Belle II FEI hadronic tagging algorithm. Signal side is reconstructed by charged lepton and hadronic part $X$ from all remaining tracks and clusters after tag side reconstruction called Rest of Event (ROE). 

The identification of inclusive $\bar{B} \to X_u \ell\bar{\nu}_{\ell}$ decays is difficult due to the abundance of CKM-favored $\bar{B} \to X_c \ell\bar{\nu}_{\ell}$ events which share a similar event topology. Two strategies for signal distinguishing are used. The first one is the study of kaon multiplicity $N(K)$ since $K^\pm$ and $K^0_S$ predominantly tag $b \to c$ transitions. In that case $N(K) = 0$ will correspond to signal $\bar{B} \to X_u \ell\bar{\nu}_{\ell}$ enhanced sample and $N(K) > 0$ to the signal depleted sample. The second one is inclusive reconstruction of $D^*$ for additionally suppression of $\bar{B} \to X_c \ell\bar{\nu}_{\ell}$ background via soft pions identifying and events with high missing mass.

A one-dimensional fit to the charged lepton energy in the $B$ meson rest frame $E^{B_{sig}}_{\ell}$ is performed in $u$-depleted sample for $\bar{B} \to X_c \ell\bar{\nu}_{\ell}$ events normalization. While for normalization of $\bar{B} \to X_u \ell\bar{\nu}_{\ell}$ events two-dimensional fit to the $E^{B_{sig}}_{\ell}$ and squared four-momentum transfer to the lepton pair $q^2$ is used. Both fits are performed with $E^{B_{sig}}_{\ell} > 1$ GeV. Therefore, the yields are unfolded and corrected for all reconstruction and selection efficiencies.

The measured ratio $|V_{ub}|/|V_{cb}|$ is extracted using two theoretical frames BLNP and GGOU for the partial decay rate of $\bar{B} \to X_u \ell\bar{\nu}_{\ell}$:

\begin{eqnarray*}
	& {\frac{\left|V_{u b}\right|}{\left|V_{c b}\right|}}^{BLNP}=(9.81 \pm 0.42_{\text {stat.}} \pm 0.38_{\text {syst.}} \\ & \pm 0.51_{\Delta \Gamma\left(B \rightarrow X_u \ell v\right)} \pm 0.20_{\Delta \Gamma\left(B \rightarrow X_c \ell v\right)}) \times 10^{-2}, \\
	& {\frac{\left|V_{u b}\right|}{\left|V_{c b}\right|}}^{GGOU}=(10.06 \pm 0.43_{\text {stat.}} \pm 0.39_{\text {syst.}} \\ 
    & \pm 0.23_{\Delta \Gamma\left(B \rightarrow X_u \ell v\right)} \pm 0.20_{\Delta \Gamma\left(B \rightarrow X_c \ell v\right)}) \times 10^{-2}.
\end{eqnarray*}

These values are in excellent agreement with world averages.

\subsection{Tagged \texorpdfstring{$B \to X_u\ell\nu$}{B -> Xu l nu} decays to determine \texorpdfstring{$|V_{ub}|$}{Vub} at Belle II}

This analysis presents measurement of $|V_{ub}|$ via inclusive charmless semileptonic $B \to X_u \ell \nu$ decays \cite{Belle-II:2025rna}. Events containing $e$ or $\mu$ and a fully reconstructed by FEI algorithm $B$ meson are selected. The analysis employs three kinematical observables: lepton energy in the signal $B$ meson rest frame $E^{(B)}_{\ell}$, the hadronic system invariant mass $M_{X}$ and squared four-momentum transfer to the lepton pair $q^2$.

Analysis is performed in a part of the phase-space to mitigate the overwhelming background emerging from CKM-favored semileptonic decays to a hadronic system containing a charm quark. Suppression of the background events is enabled by the implementation of neural network models. In order to suppress the continuum contribution, a multi layer perceptron (MLP) is trained using topological observables. Another MLP model is trained for dominant background from $B \to X_c \ell \nu$ separation. Classifier's input features exploit differences between semileptonic decays to charmed and charmless mesons such as missing mass squared $M^2_{miss}$, $p$-value of the ROE vertex fit and others. Additionally, kaon multiplicity is studied.

For precise modeling and normalization corrections, the dataset is divided into six regions based on kaon multiplicity and $b \to c$ suppression classifier output. For signal extraction two regions are used: a signal region with classifier scores above $0.87$, and a control region with scores below $0.60$, both with implemented kaon veto.

A simultaneous two-dimensional fit to $E^{(B)}_{\ell}$:$q^2$ distributions is performed in the broadest phase-space region defined by selection $E^{(B)}_{\ell} > 1$ GeV, with the most reliable theoretical predictions, The obtained value of partial branching fraction is:
\begin{equation*}
    \Delta \mathcal{B}\left(B \rightarrow X_u \ell v\right)=(1.54 \pm 0.08 \pm 0.12) \times 10^{-3}.
\end{equation*}

From this branching fraction, the value of $|V_{ub}|$ obtained in the GGOU framework is:

\begin{equation*}
    \left|V_{u b}\right|=\left(4.01 \pm 0.11 \pm 0.16_{-0.08}^{+0.07}\right) \times 10^{-3}.
\end{equation*}

Measured value is compatible with the average. 

\subsection{\texorpdfstring{$|V_{cb}|$}{Vcb} measurement from untagged \texorpdfstring{$B \to D \ell \nu$}{B -> D l nu}}
This analysis presents the exclusive $|V_{cb}|$ determination from untagged semileptonic decays on 365 $\text{fb}^{-1}$ of Belle II data \cite{Belle-II:2025rnaa}. For the measurement of $|V_{cb}|$, the decay $B \to D \ell \nu$ has received less attention than $B \to D^* \ell \nu$ due to its smaller branching fraction and significant background. However, these disadvantages can be mitigated with the large data samples.  

Signal $B$ meson candidates are reconstructed in two modes $B^0 \rightarrow D^- (\to K^- \pi^+ \pi^+) \ell \nu$ and $B^+ \rightarrow \bar{D}^0 (\to K^- \pi^+) \ell \nu$, where $\ell$ denotes $e$ or $\mu$. The second $B$ meson from $\Upsilon(4S)$ decay is not reconstructed explicitly. Instead, an inclusive reconstruction of the unobserved neutrino momentum is used. This method allows to use recoil variable $w = v_B \times v_D$, where $v_B$ and $v_B$ -- the four-velocities of the $B$ and $D$ mesons. Another variable is $cos{\theta_{BY}}$, defined by Eq. (\ref{cosine}) with $Y = D\ell$. 

Total branching fractions of each mode are measured with fit on $\mathrm{\cos{\theta_{BY}}}$ in bins of $w$. The obtained result for both mods: 
\begin{eqnarray*}
    &\mathcal{B}\left(B^0 \rightarrow D^{-} \ell^{+} \nu_{\ell}\right)=  2.06 \pm 0.05_\text{stat.} \pm 0.10_\text {syst.} \%,\\ 
    &\mathcal{B}\left(B^{+} \rightarrow \bar{D}^0 \ell^{+} \nu_{\ell}\right)= 2.31 \pm 0.04_\text{stat.} \pm 0.09_\text {syst.} \%.
\end{eqnarray*}

Finally, $|V_{cb}|$ measurement obtained with Bourrely-Caprini-Lellouch (BCL) parameterization is:

\begin{equation*}
    |V_{cb}|_{BCL} = (39.2 \pm 0.4_{\text{stat.}} \pm 0.6_{\text{syst.}} \pm 0.5_{\text{th.}}) \times 10^{-3}.
\end{equation*}

This measurement is consistent with other recent exclusive measurements. Moreover, this analysis provides the most precise determination of $|V_{cb}|$ with semileptonic $B \to D \ell \nu_{\ell}$ decay data.

\section{Summary}

We reported recent measurements of the CKM matrix elements including both untagged and tagged analyses with exclusive or inclusive final states and LFU tests using Belle or Belle II dataset. As was demonstrated, modern analyses show strong experimental performance and are competitive with previous measurements even with limited statistics. Belle II experiment is currently collecting more data, therefore, we hope to obtain more accurate measurements and contribute to the fundamental tests of the SM and beyond.

This work was funded by the Russian Science Foundation (project No. 25-12-00160).

\nocite{*}


\end{document}